\documentclass[twocolumn,showpacs,preprintnumbers,superscriptaddress,amsmath,amssymb]{revtex4}
\usepackage{graphicx}
\usepackage{dcolumn}
\usepackage{bm}

\begin{document}

\title{\bf Simple Analytic Models of Gravitational Collapse  } 

\author{R. J. Adler} %
\affiliation{Hansen Laboratory for Experimental Physics, Stanford
University, Stanford, CA 94309, USA}
\author{J. D. Bjorken} %
\affiliation{Stanford Linear Accelerator Center, Stanford
University, Stanford, CA 94309, USA}
\author{P. Chen} %
\affiliation{Stanford Linear Accelerator Center, Stanford
University, Stanford, CA 94309, USA}
\author{J. S. Liu} %
\affiliation{Department of Physics, Stanford University,
Stanford, CA } %

\date{\today}

\begin{abstract}
Most general relativity textbooks devote considerable space to the
simplest example of a black hole containing a singularity, the
Schwarzschild geometry. However only a few discuss the dynamical
process of gravitational collapse, by which black holes and
singularities form. We present here two types of analytic models
for this process, which we believe are the simplest available; the
first involves collapsing spherical shells of light, analyzed
mainly in Eddington-Finkelstein coordinates; the second involves
collapsing spheres filled with a perfect fluid, analyzed mainly in
Painleve-Gullstrand coordinates. Our main goal is pedagogical
simplicity and algebraic completeness, but we also present some
results that we believe are new, such as the collapse of a light
shell in Kruskal-Szekeres coordinates.
\end{abstract}



\maketitle

\section{Introduction}

Black holes and singularities are certainly some of the most
peculiar and interesting features of general
relativity\cite{Davies}. From the perspective of fundamental
theory their properties are far from completely understood,
especially their relation to quantum theory\cite{Fraser}; from the
perspective of astrophysics and observational astronomy they seem
more and more to play a central role in the physics of objects
ranging from stellar to cosmological
size\cite{Misner-Thorne-Wheeler}.

Most general relativity textbooks devote considerable space to
discussing the simplest example of a black hole containing a
singularity, the Schwarzschild
geometry\cite{Misner-Thorne-Wheeler,Adler,Landau,Ohanian,Kenyon,
Weinberg,Rindler,Schutz,Wald,D'Inverno}, and some also discuss
Kerr's generalization to a rotating system, although only a few
derive it\cite{Adler,Schiffer}. Most also explain, at least
qualitatively, why black holes should exist in the real world,
based on the instability of neutron stars of more than about 4
solar masses; even neutron degeneracy pressure is not sufficient
to provide stability. A few texts also discuss Schwarzschild's
interior solution for a constant density
star\cite{Adler,Weinberg}, which clearly illustrates some of the
same instability features as more realistic neutron star models.
But the actual dynamical process of collapse, whereby a massive
body becomes a black hole, is a more complex dynamical problem,
and is often either neglected or treated heuristically and
qualitatively.

For a brief history of gravitational collapse see
reference\cite{Hillman}. The study of gravitational collapse
dynamics began with the classic seminal work of Oppenheimer and
Snyder\cite{Openheimer}, who set up the equations and gave a
semi-quantitative discussion of the collapse of a spherical
non-radiating star. They also found a complete analytic solution
for the case of a uniform perfect fluid of zero pressure - now
widely referred to as "dust." This example is less artificial than
it might appear since the effect of pressure in a more realistic
treatment turns out to be not very critical; but the treatment by
Oppenheimer and Snyder involves some awkward algebra in its
treatment of the boundary between the interior and exterior of the
dust star.

Some textbooks discuss collapse much like Oppenheimer and Snyder,
notably Landau and Lifshitz\cite{Landau} and Misner Thorne and
Wheeler\cite{Misner-Thorne-Wheeler}. In particular these texts use
a Novikov type coordinate system, in which the dust is co-moving
in a bound orbit\cite{Novikov}. This coordinate system is
synchronous and co-moving, and is wonderfully simple conceptually,
but almost equally awful algebraically. Due to the algebraic
difficulty, we believe, this is not the simplest way to treat the
problem. Most other texts give a heuristic but convincing analysis
by comparing the motion of a dust particle just outside a
collapsing dust sphere with the motion of the surface; matching
this motion is equivalent to matching the exterior and interior
geometries, which is a rather obvious but important
theorem\cite{Harrison}.

What we have tried to do in the present paper is to analyze
dynamical collapse in as simple a way as possible by our choice of
coordinates; we use
Eddington-Finkelstein\cite{Eddington,Finkelstein} type coordinates
for infalling incoherent light or "null dust," and
Painleve-Gullstrand\cite{Painleve,Gullstrand} type coordinates for
perfect fluids, including dust. These coordinates seem almost
magically well suited to this purpose, as we hope will become
apparent in sections 2 and 5. We use zero energy fluid systems in
contrast to negative energy bound systems as used in references
[3] and [5], largely because the algebra is much
simpler\cite{Adler}. This choice seems to us well justified
because the zero and nonzero energy systems have the same
qualitative behavior as they approach the black hole state as seen
from outside in terms of standard Schwarzschild time, and have
exactly the same behavior as they approach the singularity as seen
in terms of the proper time of a falling observer. Moreover the
zero energy case does not have the awkward conceptual problem of
the motion before the maximum size is reached, since the maximum
size occurs at an infinite time in the past. We therefore believe
the zero energy case is more illustrative and economical of
effort.

Almost needless to say we have treated only spherically symmetric
collapsing systems. The Birkhoff theorem guarantees that a
spherically symmetric system does not emit gravitational
radiation; non-spherically symmetric collapse generally entails
gravitational radiation, making the problem vastly more complex,
though also more interesting.

To achieve our goal of simple textbook type examples of collapse,
we must pay a price, which is some degree of artificiality. The
main artificiality is the spherical symmetry; also the in-falling
shells of light or other massless material of sections 2 and 3 are
not likely to be found in nature; similarly we do not expect to
find the perfect fluid sphere surrounded by an elastic shell of
section 5, or the zero pressure perfect fluid of sections 6 and 7.

Since our purpose in the present paper is overtly pedagogical,
much of what we obtain here is known and available somewhere in
the vast research literature, although obtained with different
techniques and different coordinates. However we believe our
treatment is the simplest, both conceptually and algebraically,
mainly due to our coordinate choice. We do not know of any use of
Painleve-Gullstrand coordinates in the present context, although
they are becoming more widely used\cite{Parikh-Wilczek,Parikh}; as
will be seen in section 6 these coordinates combine a
Schwarzschild type radial coordinate and a
Friedmann-Robertson-Walker (FRW) type cosmological time coordinate
to give a natural interpolating system for describing both the
interior and exterior of a fluid sphere. While the
Eddington-Finkelstein and Painleve-Gullstrand coordinates serve
their mathematical purpose quite well, they have a number of
drawbacks such as a ``time" marker that is not well-behaved
everywhere in spacetime, and they are not synchronous; we have
therefore also treated the collapse of a thin shell of light by
transforming to Kruskal-Szekeres coordinates, which are
conformally flat and thus display the causal structure of the
collapse process quite clearly\cite{Kruskal,Szekeres}. In
Kruskal-Szekeres coordinates one has a nice picture for
distinguishing real world black holes from "eternal" black holes,
with their associated wormholes and white hole segments, none of
which may be expected to actually exist in our universe.

We use throughout this paper a rather novel technique for handling
the boundary conditions between the fluid and the exterior, which
we believe is physically clear.

Since the pioneering work of Oppenheimer and Snyder there has been
an enormous amount of work on gravitational collapse and related
topics, most directed to research rather than pedagogy. The
present pedagogical paper originated from several research
problems, one involving a formation mechanism for singularity-free
black holes filled with heavy vacuum, and another involving a
finite version of the standard or ``concordance" cosmological
model. For these we needed analytic collapse models that were
algebraically simpler and more explicit than we found in the
literature. Our hope is that our methods and models might also be
of use to students and teachers in illustrating more standard
gravitational collapse.

This paper is organized as follows: the first part (sections 2 to
4) deals with collapsing shells of incoherent light or null dust,
which is entirely characterized by its null 4-velocity and energy
density; the second part (sections 6 and 7) deals with spheres of
perfect fluid, including some with nonzero pressure and some with
non-uniform density. In section 2 we use Eddington-Finkelstein
coordinates to obtain the metric for the collapse of a thin light
shell onto a black hole to form a larger black hole, and the
collapse of a thin light shell to form a black hole ab initio. The
ab initio collapse is amusing in that it is almost certainly the
simplest complete scenario for the formation of a black hole,
albeit it a rather artificial one. In section 3 we use the results
of section 2 to obtain the metric for the collapse of a thick
shell of light, layer by layer. The density profile of the shell
is largely arbitrary, and we give one specific example. In section
4 we transform the metric for the thin light shell collapse to
conformably flat Kruskal-Szekeres coordinates in order to display
the causal structure of the geometry in a clear way - nearly
equivalent to a Penrose diagram; as already noted we use these
coordinates to emphasize the difference between black holes formed
by collapse and eternal Schwarzschild black holes. In section 5 we
present our way of doing the classic problem of collapse of a
uniform fluid sphere, using Painleve-Gullstrand coordinates for
maximum simplicity. We also obtain an explicit solution for a
nonzero pressure fluid with a linear equation of state:
$p=\alpha\rho$; to balance the pressure gradient force at the
surface we use the artifice of a thin elastic shell with
tangential pressure instead of the slowly decreasing radial
pressure of a real stellar system; our model is essentially a
balloon with uniform internal density and pressure. The stress
energy tensor of the surface is described by the radius of the
sphere, its energy density, and the parameter $\alpha$; for
$\alpha=0$ we recover the standard results for the dust ball. In
section 6 we deal with zero pressure non-uniform dust spheres,
with the density characterized by a largely arbitrary function;
again using Painleve-Gullstrand coordinates we build up the system
layer by layer, to obtain the standard results of Oppenheimer and
Snyder and Landau and Lifshitz etc. We do not allow the dust
layers to cross, which could give rise to infinite densities and
superficial singularities\cite{Singh}. We give one specific
example of the metric of a non-uniform sphere; the special case of
uniform density agrees with the analysis of section 5. Since this
paper is mainly pedagogical we have throughout sacrificed brevity
and included considerable algebraic detail; only elementary
mathematical methods are used.

It is worth mentioning that some simple and fairly obvious
extensions of the present work can be made. First, by a reversal
of time the collapse of light shells can be viewed as outgoing
radiation, so we may easily obtain results like those of Vaidya
for the geometry of a radiating body\cite{Vaidya}. Similarly the
collapse of the perfect fluid can be time reversed to yield the
metric for a black hole emitting matter, that is a type of white
hole. Also it is straightforward to include a cosmological
constant term in the equations to obtain the metric for a
collapsing system in an exterior Schwarzschild de Sitter geometry;
due to its somewhat more cumbersome algebra we have not included
this in the present paper, and leave it as an exercise. (See
reference\cite{Lake}.) We will say more about further extensions
and applications in section 8.

\section{Thin Light Shells}

The motion of light in Schwarzschild geometry is algebraically
awkward at and inside the Schwarzschild radius when Schwarzschild
coordinates are used, but it becomes quite simple in
Eddington-Finkelstein (EF) coordinates. In this section we use EF
coordinates to study the motion of light, the growth of a black
hole by absorption of light, and the ab initio formation of a
black hole by a thin shell of light. The last process is most
likely the simplest model for the formation of a black hole by
gravitational collapse.

The Schwarzschild metric in Schwarzschild coordinates is, with
$c=1$,
\begin{eqnarray}
ds^2 = (1+u)dt_s^2 - \frac{dr^2}{1+u} - r^2 d\Omega^2, \quad
\label{Schwarzschild}
\end{eqnarray}
where $u = -R/r$ and $d\Omega^2=d\theta^2 +\sin^2\theta d\phi^2$.
The black hole surface is a sphere at the Schwarzschild radius
(twice the geometric mass) $R=2m=2GM$. It is both an infinite
redshift surface where $g_{00}=0$ and a null surface or
one-way-membrane. In terms of the Schwarzschild time coordinate
both light and particles take an infinite time to reach the black
hole surface from the exterior. Specifically, for light falling
radially from $r_i$
\begin{eqnarray}
\Delta t_s = r_i - r + R \ln\Big(\frac{r_i-R}{r-R}\Big).
\end{eqnarray}
Thus light never reaches the surface but approaches asymptotically
with a characteristic time $R$. It is for this reason that we
introduce the EF time coordinate, in terms of which the surface is
reached in a finite time.
    The EF time coordinate is obtained from the Schwarzschild time by
a radially dependent shift
\begin{eqnarray}
t_s = t + g(r). \label{EFt}
\end{eqnarray}
This leads to the EF form of metric,
\begin{eqnarray}
ds^2 = dt^2 -dr^2 - r^2 d\Omega^2 + u(dt \pm dr)^2, \label{EF}
\end{eqnarray}
provided that we choose the transformation function $g$ to obey
\begin{eqnarray}
\frac{dg}{dr} = \pm \frac{u}{1+u}.
\end{eqnarray}
The solution to this is
\begin{eqnarray}
g = \mp R\ln\Big(\frac{r}{R}-1\Big),
\end{eqnarray}
which has an infinite stretch at $r=R$. Henceforth we refer to the
coordinates and metric form in Eq.(\ref{EF}) as EF for any
function $u(r,t)$.

\begin{figure}[ht]
\includegraphics[scale=0.5]{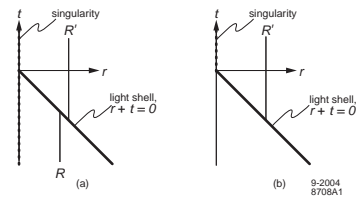}
\caption{\footnotesize{In (a) a thin light shell falls into a
black hole to produce a larger black hole. In (b) a thin light
shell with flat Minkowski interior collapses to form a black
hole.}} \label{ThinLightShellEF}
\end{figure}

The EF form is extremely convenient because radially infalling
light behaves quite simply if the plus sign in Eq.(\ref{EF}) is
chosen. For such light we set the line element equal to zero, to
obtain
\begin{eqnarray}
 ds^2 &=& dt^2 - dr^2 + u(dt+dr)^2  \nonumber \\
      &=& (dt+dr)(dt - dr + udt + udr) = 0.
\end{eqnarray}
so that
\begin{eqnarray}
dt + dr = 0, \quad t+r = {\rm const},  & \quad \textrm{infalling light}, \\
dr/dt = (1+u)/(1-u), & \quad \textrm{outgoing light}.
\label{EFLight}
\end{eqnarray}
Thus the path of infalling light is independent of the metric
function $u$, and is the same as in flat space. See
Fig.~\ref{ThinLightShellEF}. Conversely, outgoing light ``stalls''
where $u=-1$ and $g_{00}=0$, that is at the infinite redshift
surface.

We will soon need the Einstein tensor for the EF metric form
Eq.(\ref{EF}). Its nonzero components are easily calculated to be
\begin{eqnarray}
& & G^0_{\;0} = \frac{u^{\prime}}{r}+\frac{u}{r^2}, \quad
G^1_{\;0} = -\frac{\dot{u}}{r}, \quad G^0_{\;1} =
\frac{\dot{u}}{r}, \nonumber \\
& & G^1_{\;1} = \frac{u^{\prime}}{r}+\frac{u}{r^2}-\frac{2\dot{u}}{r}, \nonumber \\
& & G^2_{\;2} = G^3_{\;3} =
\frac{u^{\prime}-\dot{u}}{r}+\frac{\ddot{u}+u^{\prime\prime}-2\dot{u}^{\prime}}{2}.
\label{EFG}
\end{eqnarray}
Here a dot denotes a time derivative and a prime denotes a radial
derivative. We define the stress energy tensor in terms of the
Einstein tensor by using the field equations
\begin{equation}
G^{\alpha}_{\;\beta} \equiv -8\pi GT^{\alpha}_{\;\beta}.
\label{EinsteinEquation}
\end{equation}
For pure Schwarzschild geometry the Einstein tensor is everywhere
zero, as is easily verified.

We now study the formation of a black hole by a single thin shell
of light, and in the next section we will consider a thick shell
of light. For both purposes we begin with a thin shell of light
falling into a Schwarzschild black hole with radius $R$  to form a
black hole with radius $R^{\prime}$  as shown in
Fig.~\ref{ThinLightShellEF}a; the light shell obeys $r+t=0$
everywhere. We may write the metric in all of spacetime using a
step function $\Theta$ and its complement $\tilde{\Theta} =
1-\Theta$ , as
\begin{eqnarray}
u = -\frac{R}{r}\tilde{\Theta}(r+t) -
\frac{R^{\prime}}{r}\Theta(r+t),
\end{eqnarray}
which is of course time dependent. Substituting this into
Eq.(\ref{PGG}) and Eq.(\ref{EinsteinEquation}) we obtain the
stress-energy tensor for the thin light shell, which is singular
due to the step function,
\begin{eqnarray}
T^{\alpha}_{\;\beta} = \frac{R^{\prime}-R}{8\pi Gr^2} \delta(r+t)
k^{\alpha} k_{\beta}, \quad
\label{ThinLightShellEFT}
\end{eqnarray}
where $k^{\alpha}=(1,-1,0,0)$ and $k_{\beta}=(1,1,0,0)$. The null
vector $k^{\alpha}$ corresponds to infalling light. From the
energy density $T^0_{\;0}$ we may calculate the energy or
effective mass of the light shell by going to a time in the
distant past when the shell was in asymptotically flat space,
\begin{eqnarray}
M_s = \int 4\pi r^2 T^0_{\;0} dr = \frac{R^{\prime}-R}{2G} =
M^{\prime}-M.
\end{eqnarray}
This verifies that the mass of the initial black hole plus the
energy or effective mass of the light shell equals the mass of the
final black hole.

The above results hold for the special case when the initial black
hole is replaced by flat space, or $R=0$.  Note that everything
above is consistent with the interior of a spherical thin light
shell being flat Minkowski space. This represents the formation of
a black hole by the gravitational collapse of a single thin light
shell (or other massless material), as shown in figure 1b; this
would seem to be the simplest complete example of gravitational
collapse.

\section{Thick Light Shells}

\begin{figure}[ht]
\includegraphics[scale=0.5]{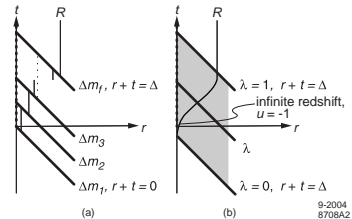}
\caption{\footnotesize{In (a) a discrete sequence of light shells
forms a black hole; (b) shows the continuous version of the same
process.}} \label{ThickLightShellEF}
\end{figure}

    Our results from the preceding section may be used to construct a
model for the collapse of a thick light shell, layer by layer. The
initial state is a sequence of concentric thin shells with a flat
Minkowski interior, as shown in Fig.~\ref{ThickLightShellEF}a.
Each region of spacetime has a Schwarzschild geometry as discussed
in the previous section. The innermost shell with energy $\Delta
m_1$ collapses to form a black hole of mass $m_1 =\Delta m_1$,
followed by others to form intermediate black holes of mass $m_j =
\Delta m_1 + \cdots + \Delta m_j$, ending with a final black hole
with mass $m_f$ and Schwarzschild radius $R = 2m_f$. Between the
shells the geometry is given by Eq.(\ref{EF}), with a metric
function
\begin{eqnarray}
u = -2m_j / r. \label{DiscreteU}
\end{eqnarray}
The evolving infinite redshift surface, defined by $u=-1$, is the
zigzag line in Fig.~\ref{ThickLightShellEF}a.

    The continuous analog of the sequence is a family of concentric
shells with a continuous label $\lambda$, which we choose to run
from 0 to 1. The parameter $\lambda$ plays the same role as the
discrete label $j$: that is the energy or effective mass inside
the $\lambda$ shell is denoted by $m(\lambda)$. As in
Eq.(\ref{DiscreteU}) the metric function in the light shell is
\begin{eqnarray}
u = -2m(\lambda) / r.
\end{eqnarray}
The evolving infinite redshift surface is the smooth line in
Fig.~\ref{ThickLightShellEF}b.

    A labeling scheme, which is convenient for both the present light
shell and the fluid systems to be discussed later, is to take the
energy inside the $\lambda$ shell to be proportional to $\lambda$,
\begin{eqnarray}
m(\lambda) = \lambda m_f, \quad m_f = \textrm{final mass}.
\end{eqnarray}
The time at which the $\lambda$ shell reaches the center may be
chosen almost arbitrarily as a function of $\lambda$, subject to
the constraints that it be 0 for the innermost shell $\lambda=0$
and increase monotonically to the final time $\Delta$ for the
outermost shell $\lambda=1$, as shown in
Fig.~\ref{ThickLightShellEF}. We denote this function as
$F^{-1}(\lambda) \Delta$, for reasons that will become apparent.
The equation of motion for the shell $\lambda$ is thus
\begin{eqnarray}
r + t = F^{-1}(\lambda) \Delta.
\end{eqnarray}
From its definition $F^{-1}$ must have an inverse, so we may
invert this to obtain
\begin{eqnarray}
\lambda = F\Big(\frac{r+t}{\Delta}\Big), \quad m(\lambda) = m_f
\lambda = m_f F,  \label{3.5}
\end{eqnarray}
with $F(0)=0$ and $F(1)=1$. Thus $F$ serves as a density profile
function, and the metric in the interior of the light shell may be
written as
\begin{eqnarray}
u = -\frac{2m_f}{r} F\Big(\frac{r+t}{\Delta}\Big) = -\frac{RF}{r}.
\label{ContinuousU}
\end{eqnarray}
The infinite redshift surface, defined by $u=-1$, is thus
determined by
\begin{eqnarray}
r = R \, F\Big(\frac{r+t}{\Delta}\Big).
\label{ThickLightShellInfiniteRedshiftSurface}
\end{eqnarray}
Inverting this we obtain for the infinite redshift surface
\begin{eqnarray}
t = \Delta F^{-1}(r/R) - r.
\end{eqnarray}
This implies that if the total duration of collapse $\Delta$ is
sufficiently large then $t$ must be a positive monotonic function
of $r$.

    Using the metric function Eq.(\ref{ContinuousU}) we may calculate the Einstein
tensor and the stress-energy tensor from Eq.(\ref{EFG}) and
Eq.(\ref{EinsteinEquation}). This gives for the light shell,
\begin{eqnarray}
T^{\alpha}_{\;\beta} = \frac{RF^{\prime}}{8\pi Gr^2 \Delta}
k^{\alpha} k_{\beta}, \quad \textrm{light shell},
\end{eqnarray}
where $F^{\prime}$ denotes the derivative of $F$ with respect to
its argument, and $k^{\alpha}$ is the null vector defined in
Eq.(\ref{ThinLightShellEFT}). From this we obtain the energy
density and the total energy of the light shell as
\begin{eqnarray}
& & \rho  =  T^0_{\;0} =\frac{RF^{\prime}}{8\pi Gr^2 \Delta},
\nonumber \\
& &  M_s  =  \int 4\pi r^2 \rho dr = \frac{m_f F(1)}{G} =
\frac{m_f}{G},
\end{eqnarray}
where the integral is again done in the asymptotically flat
spacetime of the distant past, verifying that the energy of the
light shell is equal to the final black hole mass.

    Our approach to boundaries and boundary conditions is somewhat
unorthodox; we do not impose boundary conditions per se on the
metric function between regions of spacetime. Instead we use the
above solutions (in vacuum and within the light shell) to
calculate the stress-energy tensor, which is defined by the field
equations and Eq.(\ref{EFG}).  The resulting stress energy tensor
must be zero in vacuum, correctly represent the stress energy
within the light shell, and describe the stress on the boundaries.
If it does, the solution makes physical sense. For the region near
the inner and outer boundaries of the light shell we may write the
metric function as
\begin{eqnarray}
u = \left\{\begin{array}{ll}
-(RF/r) \Theta(r+t), & \textrm{inner surface}, \\
-(RF/r)\tilde{\Theta}(r+t-\Delta), & \textrm{outer surface}.
\end{array} \right.
\end{eqnarray}
Calculating the stress energy tensor from Eq.(\ref{EFG}) and
Eq.(\ref{EinsteinEquation}) we find that there is no singular
shell at these boundaries since the relevant singular derivatives
cancel; thus the stress-energy tensor associated with the
boundaries is zero and the solution is thus physically reasonable.

    As a specific example let us take the density profile function to be
linear,
\begin{eqnarray}
F\Big(\frac{r+t}{\Delta}\Big) = \frac{r+t}{\Delta}, \quad
m(\lambda) = m_f \Big(\frac{r+t}{\Delta}\Big).
\end{eqnarray}
This corresponds to a constant rate of energy impacting the
center. From Eq.(\ref{ContinuousU}) the metric function in the
various regions is,
\begin{eqnarray}
u = \left\{ \begin{array}{ll}
0, &  \textrm{Minkowski region},  \\
-R/r, & \textrm{Schwarzschild region},  \\
-R(r+t)/(r\Delta), & \textrm{within light shell}.
\end{array} \right.
\end{eqnarray}
The infinite redshift surface, from
Eq.(\ref{ThickLightShellInfiniteRedshiftSurface}), is the linear
function
\begin{eqnarray}
t = \Big(\frac{\Delta}{R}-1\Big) r.
\label{ThickLightShellInfiniteRedshiftSurfaceExample}
\end{eqnarray}
The slope of this is positive for $\Delta/R > 1$, that is when the
light shell thickness is greater than its Schwarzschild radius.
Within the light shell the energy density is, from Eq.(\ref{EFG})
and Eq.(\ref{EinsteinEquation}),
\begin{eqnarray}
T^0_{\;0} = \rho = \frac{R}{8\pi Gr^2 \Delta}, \quad \textrm{light
shell interior,}
\end{eqnarray}
which is independent of time.

\begin{figure}[ht]
\includegraphics[scale=0.5]{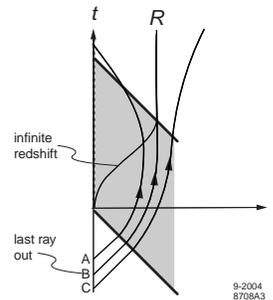}
\caption{\footnotesize{Some ``outgoing" light rays in the thick
light shell collapse.  The last ray out (B) hovers at the
Schwarzschild radius and defines a horizon.}}
\label{ThickLightShellEFHorizon}
\end{figure}

In Fig.~\ref{ThickLightShellEFHorizon} we show a qualitative
sketch of some outgoing light rays for a fairly general light
shell. The sketch is made by noting that, from Eq.(\ref{EFLight}):
the slope of outgoing rays is 1 in the Minkowski region and at
large distances it approaches 1; the slope is 0 along the infinite
redshift surface; finally, the slope is -1 at the center of the
black hole. There is a last-ray-out emitted from the origin to
infinity, after which all outgoing rays (as well as particles) are
trapped within the surface at $R$ and eventually fall into the
singularity. The surface defined by the last-ray-out is thus a
global horizon. This illustrates that the horizon and the infinite
redshift surface are quite different inside the time dependent
light shell, unlike the situation for the time independent
Schwarzschild geometry.

As a further application of our methods we note that our results
can be reversed in time to describe radiation being emitted by a
spherically symmetric system such as a star or black hole
\cite{Vaidya,Hawking,Gibbons,Hawking-Penrose}. To do this we use
the minus sign for the metric in Eq.(4) and turn the diagram in
Fig.2b upside down. It is then easy to modify our algebraic
results to describe any reasonable light shell density profile.
This may be useful in studying the final stages of black hole
evaporation, when the gravitational field of the radiation becomes
comparable to that of the black hole and cannot be neglected
\cite{Parikh-Wilczek}; it might help in determining if a black
hole radiates entirely away to vacuum or leaves behind a remnant
\cite{Adler-Santiago,Adler-Chen-Santiago}.

\section{Thin Light Shells in Kruskal-Szekeres Coordinates}

    The thin light shell discussed in section 2 is probably the
simplest model of gravitational collapse. However in EF
coordinates $g_{00}$ is negative for the interior of the black
hole. This means that if one is at coordinate rest
$(dr=d\theta=d\phi=0)$ then the square of the proper time interval
$ds^2$ has the opposite sign of the square of the coordinate time
interval $dt^2$, so that $t$ may not be interpreted as a good time
marker in that region of spacetime. The same is true of
Schwarzschild coordinates. Kruskal-Szekeres (KS) coordinates,
which are discussed in many texts, were developed to solve this
problem\cite{Kruskal,Szekeres,Adler}. Here we will obtain KS type
coordinates for the thin light shell collapse of section 2.

    By KS coordinates we mean a system in which the $t,r$ part of
the metric is conformal to flat Minkowski space, with no
singularities or zeroes. We first show how a metric in EF form,
with $u=u(r)$, may be put into conformal form. The metric
Eq.(\ref{EF}) may be factored as follows, with angular dependence
suppressed,
\begin{eqnarray}
ds^2 & = & dt^2 - dr^2 + u(dt+dr)^2 \nonumber \\
& = & (1+u)(dt+dr)(dt-\frac{1-u}{1+u}dr)  \nonumber \\
& = & (1+u) d(t+r) d(t-\sigma), \quad \sigma \equiv
\int\frac{1-u}{1+u}dr. \label{EFKS}
\end{eqnarray}
The quantities $t+r$ and $t-\sigma$ are termed conformal null
coordinates since the line element is zero along lines of constant
$t+r$ or $t-\sigma$; these thus represent radially moving light
rays. We transform to other null conformal coordinates by choosing
any functions $w(r+t)$ and $v(\sigma-t)$, so that
\begin{eqnarray}
dw = w^{\prime}(r+t) d(r+t), \quad dv = v^{\prime}(\sigma-t)
d(\sigma-t),
\end{eqnarray}
so the metric in terms of $w, v$ is
\begin{eqnarray}
ds^2 = -\frac{1+u}{w^{\prime} v^{\prime}} dw dv = H dw dv.
\label{KS}
\end{eqnarray}
The metric function $H$ may be considered to be a function of $r$
and $t$, or $w$ and $v$. We may also relate the null coordinates
to Lorentz-like coordinates,
\begin{eqnarray}
w = \tau + \rho, \quad v = \tau - \rho, \quad ds^2 = H(d\tau^2 -
d\rho^2).
\end{eqnarray}
Thus $\tau$ and $\rho$ may be interpreted as time and radial
coordinates provided that $H$ is positive and has no singularities
or zeros.

    Outside the Schwarzschild radius and the light shell
the function $\sigma$ and a convenient choice for the functions
$w$ and $v$ are the following
\begin{eqnarray}
& & \sigma(r) = \int \frac{r+R}{r-R} dr = r +
2R\ln\Big(\frac{r}{R}-1\Big), \quad r>R,  \nonumber \\
& & w_s = \gamma e^{a(t+r)}, v_s = \omega e^{a(\sigma-t)} = \omega
e^{a(r-t)} \Big(\frac{r}{R}-1\Big)^{2aR},   \nonumber \\
& & H_s = \frac{-1}{\gamma\omega a^2 r} e^{-2ar} R^{2aR}
(r-R)^{1-2aR}, \label{SchwarzschildKSTransformation}
\end{eqnarray}
where $a,\gamma,\omega = {\rm const.}$ This transformation is
chosen to make $H_s$ independent of time. In order that $H_s$ also
be nonsingular and have no zeros we choose $a=1/2R$, and to map
the exterior region to the $w_s>0, v_s<0$ quadrant we choose
$\gamma=1$ and $w=-1$. Then the transformation and metric
functions are explicitly
\begin{eqnarray}
w_s = e^{\frac{r+t}{2R}}, \quad v_s = e^{\frac{r-t}{2R}}
\Big(1-\frac{r}{R}\Big), \quad H_s = \frac{4R^3}{r}
e^{-\frac{r}{R}}. \quad \label{SchwarzschildKSTransformationOut}
\end{eqnarray}
Note that $w_s$ and $v_s$ are dimensionless and $H_s$ is the
square of a distance.

\begin{figure}[ht]
\includegraphics[scale=0.5]{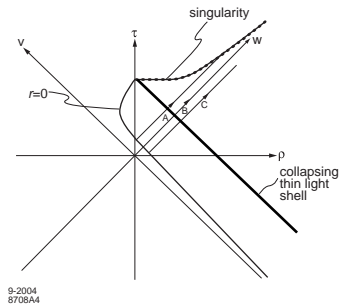}
\caption{\footnotesize{Collapse of a thin light shell in KS
coordinates, to be compared with Fig.~\ref{ThinLightShellEF}b in
EF coordinates. Only the spacetime region to the right of the line
$r=0$ has physical meaning. Compare the light rays A B C to those
in Fig.~\ref{ThickLightShellEFHorizon}.}} \label{ThinLightShellKS}
\end{figure}

For the interior of the black hole, $r<R$, and outside the light
shell (see Fig.1b) we choose the opposite sign for the logarithm
and $\gamma=\omega=1$ in Eq.(\ref{SchwarzschildKSTransformation}),
to obtain the same expression as in
Eq.(\ref{SchwarzschildKSTransformationOut}), which is thus valid
throughout the Schwarzschild geometry. The expressions
Eq.(\ref{SchwarzschildKSTransformationOut}) are similar to the
standard ones used to transform between Schwarzschild coordinates
and KS coordinates, but differ in important ways. Lines of
$r=$const. map to hyperbolae with $w_sv_s={\rm const.}$; in
particular $r=0$ corresponds to $w_sv_s=1$ and $r=R$ corresponds
to $w_sv_s=0$. The lines $t=\infty$ and $r=R$ both map to $v_s=0$,
while $t=-\infty$ maps to $w_s=0$. (Unlike the case with
Schwarzschild coordinates lines of constant $t$ do not map to rays
of $v_s/w_s={\rm const.}$!) See Fig.~\ref{ThinLightShellKS}.

There are many ways to transform from EF to KS coordinates for the
Minkowski geometry inside the light shell; of course Minkowski
geometry in EF coordinates is already in KS form, but not one that
is useful to us. We must choose a transformation that joins
continuously with the transformation
Eq.(\ref{SchwarzschildKSTransformationOut}) along the boundary
line $r+t=0$, and we also demand that the function $H$ be
continuous along that line. For that space the metric function
$u=0$, so that from Eq.(\ref{EFKS}) $\sigma=r$. Thus from
Eq.(\ref{KS}) the metric function $H$ is
\begin{eqnarray}
H_m = \frac{-1}{w^{\prime}_m(r+t) v^{\prime}_m(r-t)}.
\end{eqnarray}
Equating this with $H_s$ from
Eq.(\ref{SchwarzschildKSTransformationOut}) along $r+t=0$ we
obtain a differential equation for $v_m$
\begin{eqnarray}
v^{\prime}_m(2r) = \frac{-1}{4R^3 w^{\prime}_m(0)} r e^{r/R}.
\end{eqnarray}
That is, in terms of the argument, denoted $x$,
\begin{eqnarray}
& & v^{\prime}_m(x) = {\rm (const.)}x e^{x/2R}, \quad \nonumber \\
& & v_m(x) = {\rm (const.)}(x-2R) e^{x/2R}.
\end{eqnarray}
For continuity we choose $w_m$ to be the same as in the
Schwarzschild region Eq.(\ref{SchwarzschildKSTransformationOut}),
so with appropriate constants we arrive at
\begin{eqnarray}
& & w_m = e^{\frac{r+t}{2R}}, \nonumber \\
& & v_m = (1-\frac{r-t}{2R}) e^{\frac{r-t}{2R}}, \nonumber \\
& & H_m = 4R^3 (\frac{2}{r-t}) e^{-\frac{r}{R}}. \quad
\label{MinkowskiKS}
\end{eqnarray}
These functions are obviously equal to those in
Eq.(\ref{SchwarzschildKSTransformationOut}) along the line
$r+t=0$, as desired. Note that the price we pay for having $H_s$
in the Schwarzschild region independent of time is that $H_m$ in
the Minkowski region is dependent on time.

The nature of the transformation Eq.(\ref{MinkowskiKS}) is best
seen in terms of the mapping of some lines and points: $t=\infty$
maps to $v=0$; $t=-\infty$ maps to $w=0$; lines of constant $r+t$
map to constant $w$ lines; lines of constant $r-t$ map to constant
$v$ lines; in particular $r+t=0$ maps to $w=1$ and $r=t=0$ maps to
$w=v=1$. Finally the origin $r=0$ maps to
\begin{eqnarray}
w_m = e^{t/2R}, v_m = (1+\frac{t}{2R}) e^{-t/2R},
\end{eqnarray}
so
\begin{eqnarray}
v_m = \frac{1}{w_m}(1+\ln{w_m}).
\end{eqnarray}
Henceforth we drop the subscripts on $w$ and $v$.

Figure~\ref{ThinLightShellKS} show the complete collapse process,
with the Schwarzschild and Minkowski geometries stitched together
along the line $r+t=0$. Only the region to the right of the line
corresponding to $r=0$ has physical meaning. It is evident from
the figure that the line $v=0$ defines a horizon, and it is also
apparent that the singularity at $r=0$ does not behave like a time
independent spatial position. Note also that $\rho$ serves as a
radial marker, even though regions of the spacetime have negative
$\rho$.

\begin{figure}[ht]
\includegraphics[scale=0.5]{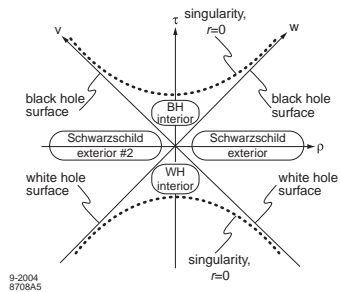}
\caption{\footnotesize{The pure or eternal Schwarzschild geometry
in KS coordinates. The entire spacetime region shown is given
physical meaning in terms of a white hole region and a second
exterior Schwarzschild region. Compare to
Fig.~\ref{ThinLightShellKS}. See for example references [3,4].}}
\label{SchwarzschildKS}
\end{figure}

Figure~\ref{ThinLightShellKS} may be compared with the "pure" or
eternal Schwarzschild geometry discussed in many textbooks
\cite{Misner-Thorne-Wheeler,Adler} and shown in KS coordinates in
Fig.~\ref{SchwarzschildKS}; this shows the "maximum analytic
extension" of the Schwarzschild solution. The region $w<0, v<0$ is
interpreted as a white hole, and is absent in figure 4; also
absent is the region $w<0, v>0$, interpreted as the "other side of
the wormhole," and part of the interior region $w>0, v<0$. This
illustrates that if the formation of the black hole by
gravitational collapse is taken into account then the
much-discussed white hole and wormhole regions are not present;
there is no reason to expect that such regions occur in nature, as
emphasized by Wheeler and many others \cite{Harrison}.

\section{Uniform Fluid Spheres}

The spherical shells of light used in the preceding sections
provide very simple models of gravitational collapse but are
rather artificial and do not approximate anything we expect to
find in nature. We now turn to a more realistic system, a sphere
of uniform perfect fluid with a linear equation of state. This
includes our approach to the special case of a dust ball, the
original \cite{Openheimer} and still a favored system for collapse
studies \cite{Misner-Thorne-Wheeler,Landau,Weinberg}. We believe
our approach is the simplest available, because the
Painleve-Gullstrand (PG) coordinate system is remarkably well
suited to the task \cite{Painleve,Gullstrand}.

Some of the mathematical techniques we used for light shells, such
as use of concentric layers and the handling of the boundary
conditions, will also prove useful for fluid spheres.

The fluid collapse involves only two spacetime regions, the fluid
interior, and the Schwarzschild exterior. A crucial step is to
find a metric form (the PG form) which describes both regions
simply, and in which the motion of the fluid is simple.

The spacetime geometry corresponding to a uniform fluid is well
known from cosmology; it is described by the
Friedmann-Robertson-Walker metric in co-moving coordinates, which
covers all of spacetime from the big bang onwards [3,4]. For
simplicity we consider the spatially flat case, that is $k=0$, for
which the metric is
\begin{eqnarray}
ds^2 = dt^2 - a(t)^2 (dr_c^2 + r_c^2 d\Omega^2). \label{FRW}
\end{eqnarray}
(This corresponds to zero energy collapsing system.) The co-moving
radial coordinate $r_c$ is dimensionless, while the scale function
$a(t)$ is a solution of the cosmological equations, with the
dimension of a length. To describe a finite sphere we truncate the
radial coordinate at some value. If the fluid has a linear
equation of state, with pressure and density obeying
$p=\alpha\rho$, then the scale function is a power of $t$,
specifically
\begin{eqnarray}
a(t) = A t^n, \quad A={\rm const.}, \quad n=\frac{2}{3(\alpha+1)}.
\label{FRWSolution}
\end{eqnarray}
In particular dust (or ``cold matter") has negligible pressure so
$\alpha=0$ and $n=2/3$, while radiation (or ``hot matter") has
$\alpha=1/3$ and $n=1/2$. This is the range of normal matter. In
the cosmological scenario time runs from the big bang at $t=0$ to
infinity, but in the collapse scenario it will run from negative
infinity to $t=0$. That is, a very large fluid sphere in the far
distant past collapses toward zero size at $t=0$.

    To obtain the desired form for the metric inside the fluid
we introduce a new radial coordinate,
\begin{eqnarray}
r = a(t) r_c,
\end{eqnarray}
which is not co-moving, to obtain the metric
\begin{eqnarray}
ds^2 & = & [1-(r\dot{a}/a)^2]dt^2 + 2(r\dot{a}/a)drdt - dr^2 -
r^2 d\Omega^2  \nonumber \\
& = & [1-(nr/t)^2]dt^2 + 2(nr/t)drdt - dr^2 - r^2 d\Omega^2.
\nonumber \\
 \label{UniformFluidPG}
\end{eqnarray}
This contains the single metric function $nr/t$ and is distinctive
in having a cross term and $g_{rr}=-1$. It has an infinite
redshift surface where $g_{00}=0$, or $nr/t=\pm1$. The minus sign
is appropriate since we will deal with negative times.

    The empty region exterior to the fluid is described by the
Schwarzschild metric in Eq.(\ref{Schwarzschild}), which we now
write in the form
\begin{eqnarray}
ds^2 = (1-\psi^2)dt_s^2 - \frac{dr^2}{1-\psi^2} - r^2 d\Omega^2,
\quad \psi=\pm \sqrt{\frac{R}{r}}. \nonumber \\
\end{eqnarray}
To make this compatible with the metric in
Eq.(\ref{UniformFluidPG}) in the fluid we choose a new time
coordinate that makes $g_{rr}=-1$. Taking
\begin{eqnarray}
t_s = t + g(r),
\end{eqnarray}
we find that
\begin{eqnarray}
ds^2 = (1-\psi^2)dt^2 \pm 2\psi drdt - dr^2 - r^2 d\Omega^2,
\label{PG}
\end{eqnarray}
provided that $g$ obeys
\begin{eqnarray}
g^{\prime} = \pm \frac{\psi}{1-\psi^2}. \label{PGtEquation}
\end{eqnarray}
As expected this transformation involves an infinite time stretch
at the Schwarzschild radius, where $\psi^2=1$. For the
Schwarzschild region the solution to Eq.(\ref{PGtEquation}) is
\begin{eqnarray}
g = \mp R \left( 2\sqrt{\frac{r}{R}} +
\ln{\frac{\sqrt{r}-\sqrt{R}}{\sqrt{r}-\sqrt{R}}} \right).
\label{PGtSolution}
\end{eqnarray}
The metric Eq.(\ref{PG}) and the transformation
Eq.(\ref{PGtSolution}) are those obtained by Painleve and
Gullstrand \cite{Painleve,Gullstrand}. Both regions of spacetime
are now described by the metric form Eq.(\ref{PG}), with $\psi$
allowed to be a function of both $r$ and $t$; specifically
\begin{eqnarray}
\psi = \left\{ \begin{array}{ll}
nr/t,         & \quad \textrm{fluid interior}, \\
-\sqrt{R/r},  & \quad \textrm{Schwarzschild exterior}.
\end{array} \right. \label{UniformFluidpsi}
\end{eqnarray}
We choose the positive signs in Eq.(\ref{PGtEquation}) and
negative sign in Eq.(\ref{PGtSolution}) to correspond to collapse
during negative time, and refer to the metric form and coordinates
as generalized Painleve-Gullstrand or simply PG.

    The PG metric has a remarkable property that is crucial to our
analysis.  The geodesic equations for radial motion of a particle
in the metric eq(\ref{PG}) lead, after some algebra, to
\begin{eqnarray}
0 &=& \frac{d^2 t}{ds^2} + \psi^{\prime} \left[
(\frac{dt}{ds})^2-1 \right], \quad \quad
\psi^{\prime} \equiv \frac{\partial \psi}{\partial r}, \nonumber \\
1 &=& (1-\psi^2)(\frac{dt}{ds})^2 +
2\psi\frac{dt}{ds}\frac{dr}{ds} - (\frac{dr}{ds})^2 .
\label{PGGeodesicEquation}
\end{eqnarray}
One obvious solution to the first is
\begin{eqnarray}
dt/ds = 1, \quad t = s - s_0.
\end{eqnarray}
Thus coordinate time and proper time intervals are equal for such
a freely falling particle, and this holds for any metric function
$\psi$. The second equation in Eq.(\ref{PGGeodesicEquation}) now
becomes quite simple
\begin{eqnarray}
\frac{dr}{ds} = \frac{dr}{dt} = \psi.
\end{eqnarray}
In the Schwarzschild region this is the same as the classical
Newtonian equation for a radially falling test particle of zero
energy.

    We will soon need the Einstein tensor and the stress-energy
tensor. From the PG metric form it is straightforward and only
slightly tedious to calculate these.  The nonzero components of
the Einstein tensor are
\begin{eqnarray}
& & G^0_{\;0} = -\frac{2\psi\psi^{\prime}}{r} -
\frac{\psi^2}{r^2}, \quad G^1_{\;0} = \frac{2\psi\dot{\psi}}{r},
\nonumber \\
& & G^1_{\;1} = -\frac{2\psi\psi^{\prime}}{r} - \frac{\psi^2}{r^2}
 - \frac{2\dot{\psi}}{r}, \nonumber \\
& & G^2_{\;2} = G^3_{\;3} =
-\frac{\dot{\psi}+2\psi\psi^{\prime}}{r} - \dot{\psi}^{\prime} -
\psi\psi^{\prime\prime} - \psi^{\prime\,2}. \label{PGG}
\end{eqnarray}
As before we define the stress-energy tensor via the field
equations $G^{\alpha}_{\;\beta} = -8\pi GT^{\alpha}_{\;\beta}$; in
particular the energy density is
\begin{eqnarray}
 \rho = T^0_{\;0} &=& \frac{1}{8\pi G}
(\frac{2\psi\psi^{\prime}}{r} + \frac{\psi^2}{r^2}) \nonumber\\
 &=& \left\{
\begin{array}{cl}
0, & \textrm{exterior,} \\
\frac{3n^2}{8\pi Gt^2}, & \textrm{interior.}
\end{array} \right. \label{UniformFluidT00}
\end{eqnarray}
Thus the density is uniform, as expected.

\begin{figure}[ht]
\includegraphics[scale=0.5]{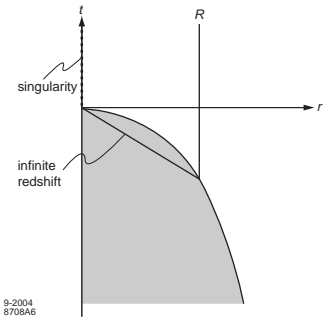}
\caption{\footnotesize{Collapse of a uniform fluid sphere to form
a black hole in PG coordinates.}} \label{UniformFluidSpherePG}
\end{figure}

We now join the spacetime regions inside and outside the fluid by
demanding that the metric function $\psi$ in
Eq.(\ref{UniformFluidpsi}) be continuous across the fluid surface
boundary. This gives
\begin{eqnarray}
r^{3/2} + \frac{\sqrt{R}}{n} t = 0. \label{UniformFluidBoundary}
\end{eqnarray}
The geodesic equation for a zero energy falling particle in the
Schwarzschild exterior region is
\begin{eqnarray}
r^{3/2} + \frac{3\sqrt{R}}{2} (t-t_0) = 0.
\label{PGGeodesicSolution}
\end{eqnarray}
This agrees with the motion of the surface
Eq.(\ref{UniformFluidBoundary}) only for the case of $n=2/3$, that
is dust. Thus a freely falling particle may hover at the falling
surface of a dust ball, as it should. However, if the pressure is
nonzero and $n<2/3$ then the surface will fall more rapidly. This
is due the stress in the surface layer, as we will see.
Figure~\ref{UniformFluidSpherePG} shows the collapse scenario in
PG coordinates.

It is straightforward to calculate the stress-energy tensor for
the fluid surface with the same technique we used in section 3. In
the vicinity of the surface the metric function $\psi$ may be
written
\begin{eqnarray}
\psi = \frac{nr}{t} \tilde{\Theta}(r^{3/2}+\frac{\sqrt{R}}{n}t) -
\sqrt{\frac{R}{r}} \Theta(r^{3/2}+\frac{\sqrt{R}}{n}t).
\label{FluidSchwarzschildpsi}
\end{eqnarray}
This leads to a singular stress-energy tensor
\begin{eqnarray}
T^2_{\;2} = T^3_{\;3} = \frac{nr}{8\pi Gt^2} (1-\frac{3n}{2}) \,
\delta\left( r - (\sqrt{R}t/n)^{2/3} \right).
\label{UniformFluidT22}
\end{eqnarray}
For a zero pressure fluid, $\alpha=0$ and $n=2/3$, this vanishes
as expected. If the pressure is not zero it represents a surface
tension, as we will now show. The surface tension of a fluid
sphere is related to its radius and pressure by $\tau=pr/2$; in
the present case this implies from Eq.(\ref{FRWSolution}) and
Eq.(\ref{UniformFluidT00}),
\begin{eqnarray}
\tau = \frac{pr}{2} = \frac{\alpha\rho r}{2} = \frac{nr}{8\pi
Gt^2}(1-\frac{3n}{2}), \label{SurfaceTension}
\end{eqnarray}
which agrees with Eq.(\ref{UniformFluidT22}). Thus the
stress-energy tensor indeed represents a surface tension that
balances the internal pressure of the fluid to keep it stable.
This is why the surface falls faster than a freely falling
particle.

In summary the collapse of a uniform fluid sphere is described by
the metric function in Eq.(\ref{PG}) and
Eq.(\ref{UniformFluidpsi}), the density in
Eq.(\ref{UniformFluidT00}), and the surface tension in
Eq.(\ref{UniformFluidT22}). Its collapse is qualitatively similar
to that of the light shells, but rather less artificial. Our use
of surface tension to stabilize the surface is a conceptually
simple substitute for the gradual pressure gradient of a more
realistic model; such tangential pressures are also mentioned by
Singh in Ref.[26] and in the references contained therein.

\section{Zero Pressure Fluid Spheres}

We next consider the collapse of a sphere filled with zero
pressure perfect fluid, that is a dust ball. This problem is
discussed by Landau and Lifshitz\cite{Landau} using Lemaitre-like
coordinates\cite{Weinstein}. For simplicity and physical clarity
we use the same technique as in section 3 for constructing thick
light shells; that is we build the dust ball layer by layer from a
sequence of thin dust shells. The layers are prohibited from
crossing to prevent superficial singularities due to the
consequent infinite density\cite{Singh}. PG coordinates will again
be used, with the metric form Eq.(\ref{PG}) containing a single
metric function $\psi(r,t)$. The Einstein tensor for this is given
in Eq.(\ref{PGG}), and we recall that in PG coordinates we may
take proper time and coordinate time intervals to be equal along
zero energy particle geodesics.

\begin{figure}[ht]
\includegraphics[scale=0.5]{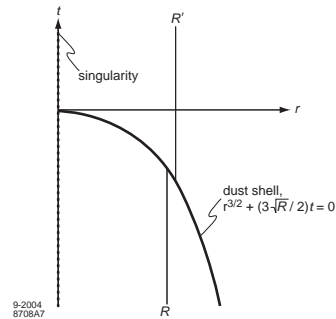}
\caption{\footnotesize{A thin shell of dust falls into a black
hole to form a larger black hole in PG coordinates. This is the
analog of Fig.~\ref{ThinLightShellEF} for light.}}
\label{DustShellPG}
\end{figure}

    In analogy with section 3 we begin with a thin shell of dust falling onto a
black hole, shown in Fig.~\ref{DustShellPG}. The shell is assumed
to be very light, so that $m$ and $m^{\prime}$ differ by a small
$\Delta m$, and the Schwarzschild radii differ by $2\Delta m$. As
in Section 5 the metric function in the two Schwarzschild regions
is
\begin{eqnarray}
\psi = \left\{ \begin{array}{ll}
-\sqrt{R/r}, & \textrm{initial Schwarzschild region}, \\
-\sqrt{R^{\prime}/r}, & \textrm{final Schwarzschild region}.
\end{array} \right.
\end{eqnarray}
The equation of the thin dust shell is that of a geodesic for a
zero energy particle in the Schwarzschild geometry, given in
Eq.(\ref{PGGeodesicSolution}). It is important that coordinate and
proper time intervals are equal along the geodesic, so that the
equation for the boundary is a relation between the coordinates
$r$ and $t$.

    To obtain the stress-energy tensor for the
dust shell we write the metric in the vicinity of the boundary as
\begin{eqnarray}
\psi = &-& \sqrt{\frac{R}{r}} \tilde\Theta(\sqrt{r^3} +
\frac{3}{2}\sqrt{R}t) \nonumber \\ &-&\sqrt{\frac{R^{\prime}}{r}}
\Theta(\sqrt{r^3}+\frac{3}{2}\sqrt{R}t).
\end{eqnarray}
where $\Theta + \tilde\Theta =1$ as before. From the Einstein
tensor in eq(ref{PGG} this leads, as in section 3, to a singular
energy density for the dust shell,
\begin{eqnarray}
\rho = T^0_{\;0} = \frac{3\sqrt{R}}{8\pi G}
\frac{\sqrt{R^{\prime}}-\sqrt{R}}{r^{3/2}}
\delta(r^{3/2}+\frac{3}{2}\sqrt{R}t).
\end{eqnarray}
From this we may calculate the mass of the shell by going to large
negative times when the shell is in nearly flat space.
\begin{eqnarray}
M_s = \int 4\pi r^2 \rho dr = \frac{\sqrt{R}}{G}
(\sqrt{R^{\prime}}-\sqrt{R}) \simeq \frac{\Delta m}{G}.
\end{eqnarray}
We thus verify that the mass of the initial black hole plus the
shell mass is equal to the mass of the final black hole.

\begin{figure}[ht]
\includegraphics[scale=0.5]{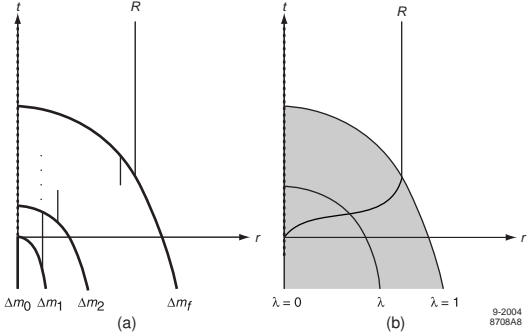}
\caption{\footnotesize{In (a) a discrete sequence of dust shells
forms a black hole, and in (b) a continuous version of the same
process forms a black hole. These are analogs of
Fig.~\ref{ThickLightShellEF} for light.}} \label{DustSpherePG}
\end{figure}

    This elementary result allows us to construct a rather general dust
ball with non-uniform density from layers of thin shells. A
discrete sequence of thin shells is shown in
Fig.~\ref{DustSpherePG}a. The initial system is a black hole of
vanishingly small mass $m_0$ with a thin shell of mass $\Delta
m_1$ collapsing onto it at $t=0$ to give a black hole of mass $m_1
= m_0 + \Delta m_1$, followed by more shells, and ending with a
final shell impacting the origin at $t=\Delta$ to give a final
black hole of mass $m_f$ and Schwarzschild radius $R=2m_f$. In
terms of the intermediate masses $m_j$ the metric function in the
regions between shells is
\begin{eqnarray}
\psi = -\sqrt{2m_j/r}, \quad \textrm{$j$-th Schwarzschild region}.
\end{eqnarray}
The infinite redshift surface where $\psi=-1$ and $g_{00}=0$ is
the zigzag line in the figure.

A continuous version of the dust shell sequence is shown in
Figure~\ref{DustSpherePG}b. The shells are labelled by a
continuous variable $\lambda$ ranging from 0 to 1, with the total
geometric mass inside a shell denoted by $m(\lambda)$;
$m(\lambda)$ is the continuum analog of $m_j$, and the metric
function in the fluid region is thus
\begin{eqnarray}
\psi = -\sqrt{2m(\lambda)/r}, \quad \textrm{fluid region}.
\label{DustSpherepsi}
\end{eqnarray}
The infinite redshift surface is where $\psi = -1$, or
\begin{eqnarray}
r = 2m(\lambda), \quad \textrm{infinite redshift surface}.
\label{DustInfiniteRedshiftSurface}
\end{eqnarray}
As in section 3 we label the shells by the total mass inside the
shell,
\begin{eqnarray}
m(\lambda) = \lambda m_f, \label{mlambda}
\end{eqnarray}
where $m_f$ is the final mass. The time at which the $\lambda$
shell impacts the center may be chosen rather arbitrarily, and we
denote it by $h(\lambda)$. For simplicity and to avoid density
singularities it is important that the shells do not cross each
other. This will be true if $h(\lambda)$ increases monotonically;
that is, outer shells impact at times later than inner shells.
Thus we choose $h(\lambda)$ to increase monotonically from 0 to
$\Delta$ as $\lambda$ runs from 0 to 1. The equation for the
$\lambda$ shell is then the free fall equation for a particle of
zero energy, with the particle reaching the origin at
$h(\lambda)$, or
\begin{eqnarray}
r^{3/2} + \frac{3}{2}\sqrt{R\lambda} [t-h(\lambda)] = 0.
\label{DustGeodesicEquation}
\end{eqnarray}
As we will see, the function $h(\lambda)$ determines the dust ball
density.

If the function $h(\lambda)$ is specified we can use
Eq.(\ref{DustInfiniteRedshiftSurface}) and Eq.(\ref{mlambda}) and
Eq.(\ref{DustGeodesicEquation}) to get a parametric expression for
the infinite redshift surface; the radius and time are given in
terms of $\lambda$ by
\begin{eqnarray}
r = R \lambda, \quad t = h(\lambda) - \frac{2}{3}R\lambda.
\end{eqnarray}
In particular for the central shell with $\lambda=0$, which
impacts the center at $t=0$, and the outer shell with $\lambda=1$,
which impacts at $t=\Delta$, we have for the infinite redshift
surface
\begin{eqnarray}
\begin{array}{ll}
r=t=0, & \textrm{central shell}, \\
r=R, \quad t=\Delta-\frac{2}{3}R,   & \textrm{outside shell}.
\end{array}
\end{eqnarray}
Thus the surface will move outward with time if $\Delta
> 2R/3$ -- that is if the mass impacts the origin
sufficiently slowly. Curiously the same expression occurred for
the light shell collapse in
Eq.(\ref{ThickLightShellInfiniteRedshiftSurfaceExample}).

    For the special case of zero total impact time, $\Delta=0$, all of the
shells impact the center at $t=0$.  This is corresponds to the
uniform density fluid sphere in the previous section, so
$h(\lambda)=0$ clearly implies uniform density. We can explicitly
see the relation between $h(\lambda)$ and the density by solving
the shell equation Eq.(\ref{DustGeodesicEquation}) for the mass
$m(\lambda)$ inside the shell $\lambda$ at a large negative time
$t=-T$, when $T>>h(\lambda)$, to get
\begin{eqnarray}
m(\lambda) = \left[ \frac{2}{9(T+h)^2} \right] r^3 \approx
\frac{2}{9T^2} r^3.
\end{eqnarray}
Thus in the distant past the mass is proportional to $r^3$,
meaning that the dust is asymptotically uniform. As time
progresses to smaller negative values the density may deviate more
and more from uniform, and may be quite non-uniform at $t=0$.

    To obtain the metric and other properties of a collapsing dust
ball we first specify a function $h$. Then the shell equation
Eq.(\ref{DustGeodesicEquation}) gives the shell parameter as a
function of position, that is $\lambda = \lambda(r,t)$. This is
consistent because we allow only one shell to pass through a given
point. With the parameter known as a function of position the
metric in the fluid is given by Eq.(\ref{DustSpherepsi}). With
this expression for the shell parameter the shell relation
Eq.(\ref{DustGeodesicEquation}) is an identity in $r,t$.

    A number of interesting properties of the dust region may be
expressed in terms of the function $h(\lambda)$. (Later we will
consider a specific $h(\lambda)$ to illustrate further.) From the
expression Eq.(\ref{DustSpherepsi}) for the metric function $\psi$
and $\lambda(r,t)$ we may calculate the Einstein tensor from
Eq.(\ref{PGG}), obtaining
\begin{eqnarray}
G^0_{\;0} = -\frac{R\lambda^{\prime}}{r^2}, \quad G^1_{\;0} =
\frac{R\dot{\lambda}}{r^2}, \quad
\lambda^{\prime}=\frac{\partial\lambda}{\partial r}, \quad
\dot{\lambda}=\frac{\partial\lambda}{\partial t}. \quad
\end{eqnarray}
For the other components we need to relate the derivatives
$\dot{\lambda}$ and $\lambda^{\prime}$ to each other, which may be
done using the shell relation Eq.(\ref{DustGeodesicEquation}).
Differentiating Eq.(\ref{DustGeodesicEquation}) we find
\begin{eqnarray}
& & \frac{\partial\sqrt{\lambda}}{\partial r} =
\frac{\sqrt{r/R}}{h-t+2\lambda dh/d\lambda}, \nonumber \\
& & \frac{\partial\sqrt{\lambda}}{\partial t} =
\frac{\sqrt{\lambda}}{h-t+2\lambda dh/d\lambda}, \nonumber \\
& & \frac{\dot{\lambda}}{\lambda^{\prime}} =
\sqrt{\frac{R\lambda}{r}}. \label{lambdaDotPrime}
\end{eqnarray}
Thus the ratio of derivatives is independent of the function $h$.
With the use of Eq.(\ref{lambdaDotPrime}) it is only slightly
tedious to show that, for any $h$,
\begin{eqnarray}
G^1_{\;1} = G^2_{\;2} = G^3_{\;3} = 0.
\end{eqnarray}
Thus the pressure terms of the stress-energy tensor are zero, as
we expect for dust. Moreover the energy density of the dust is
\begin{eqnarray}
\rho = T^0_{\;0} = \frac{R\lambda^{\prime}}{8\pi Gr^2}, \qquad
\textrm{dust region}.
\end{eqnarray}
From this we may verify that the total mass of the dust ball is
the mass of the final black hole, which we do by again going to
large negative times when the dust is in nearly flat space,
\begin{eqnarray}
M_f = \int 4\pi r^2 \rho dr = \frac{R}{2G} \int \lambda^{\prime}
dr = \frac{m_f}{G}.
\end{eqnarray}
Finally the stress energy tensor on the dust ball surface is zero
according to Eq.(\ref{SurfaceTension}) with $n=2/3$.

There are two specific examples of the function $h(\lambda)$ for
which Eq.(\ref{PGG}) is easily solved. The first example is
\begin{eqnarray}
h(\lambda) = \Delta/\sqrt{\lambda} .
\end{eqnarray}
We leave it as an exercise to the reader to show that the metric
function in the dust is then
\begin{eqnarray}
\psi =
-\frac{\sqrt{r/R}}{2\Delta}\Big[t+\sqrt{t^2+(8\Delta/3\sqrt{R})r^{3/2}}\Big].
\end{eqnarray}
In the limit of $\Delta \to 0$ this gives the same result as
Eq.(\ref{UniformFluidpsi} for a uniform dust ball. The energy
density is
\begin{eqnarray}
\rho &=& \frac{\sqrt{R}}{8\pi G\Delta
r^{3/2}}\Big[\frac{t+\sqrt{t^2+(8\Delta/3\sqrt{R})r^{3/2}}}
{\sqrt{t^2+(8\Delta/3\sqrt{R})r^{3/2}}}\Big]\nonumber\\
&\approx & 1/(2\pi G t^2), \quad \quad \textrm{for large $t$.}
\end{eqnarray}
This is a well-behaved positive function with no singularities or
zeros for negative $t$.

The second example is
\begin{eqnarray}
h(\lambda)= - D/\sqrt{\lambda} ,
 \label{DustBallFunction}
\end{eqnarray}
where $D$ is a constant. This is singular and thus quite different
from the first example, and is not represented by Fig.8b: it is
singular for $\lambda \to 0$, meaning that the central dust layer
collapsed to the center in the infinite past and the outer layer
reached the center at time $-D$. We again leave it as an exercise
to show that the metric function is
\begin{eqnarray}
\psi = (2r/3t)+\sqrt{R/r}(\Delta/t) ,
\end{eqnarray}
and that the density is
\begin{eqnarray}
\rho = D\sqrt{R}/(4\pi Gt^2r^{3/2}) .
\end{eqnarray}
The singularity at $r=0$ is mild in the sense that the mass inside
a sphere goes like $r^{3/2}$. As $t\to 0$ the density approaches
infinity as expected. Thus this example represents a dust ball
with a mildly singular central density in the distant past, which
grows stronger with time until complete collapse, and represents a
rather realistic and amusing system \cite{Liu}.

Finally, we briefly note some properties of light rays in the
collapsing dust ball. From the PG metric Eq.(\ref{PG}) we may
write the coordinate velocity of light as
\begin{eqnarray}
v_c = \frac{dr}{dt} = \left\{ \begin{array}{cl}
1+\psi,  & \textrm{outgoing}, \\
-1+\psi, & \textrm{infalling}.
\end{array} \right. \label{PGLight}
\end{eqnarray}
We generally expect that $\psi$ will be zero at $r=0$, except near
$t=0$, so $v_c=\pm 1$ there; note that this is not true for the
second example considered above. Also, $v_c=\pm 1$ for
asymptotically large distances. On the infinite redshift surface
$\psi = -1$ so that $v_c=0$ for outgoing light and $v_c=-2$ for
infalling light. For the center in the Schwarzschild region $v_c=
-\infty$. This allows us to make a rough qualitative sketch of
some light rays in Fig.9. (For $\Delta =0$ Eq.(\ref{PGLight}) may
be solved exactly.) Note that, as in the case of the thick light
shell in Section 3, there is a last light ray out, whose path
defines a global horizon inside of which neither particles nor
photons may escape to infinity.

\begin{figure}[ht]
\footnotesize{
\includegraphics[scale=0.5]{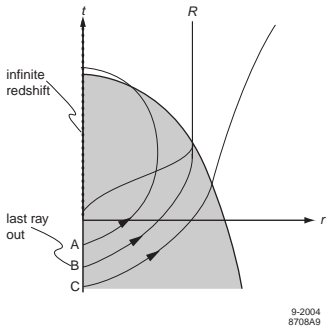}
\caption{Some ``outgoing" light rays in the collapsing dust ball.
No ray can escape from the center after the last ray out (B).
Compare to Fig.~{ThickLightShellEFHorizon} for light shell
collapse.}} \label{DustSpherePGHorizon}
\end{figure}

\section{Summary and Further Study}

In this paper we have tried to present a simple introduction to
the dynamics of gravitational collapse, which we hope can provide
a bridge between basic textbook general relativity and numerous
research topics of current interest. The light shells in EF
coordinates and the fluid spheres in PG coordinates provide the
simplest way we know of to study collapse. Some simple but amusing
extensions of these models can be made, as already mentioned: one
may reverse the time and study the emission of light and matter
from white holes or other spherically symmetric objects, and it is
easy to add a cosmological constant in the fluid collapse.

The present research literature is dense with studies involving
collapse, ranging from fundamental theory to astrophysical
applications. A sampling of recent topics from the Los Alamos
archives xxx.lanl.gov includes the following, which we paraphrase:

$\bullet$ Entropy in collapse to a black hole - where does the
information go?

$\bullet$ Unitarity - is collapse consistent with unitary quantum
evolution?

$\bullet$ Cosmic censorship - are all singularities surrounded by
a horizon?

$\bullet$ Collapse in context of string theory, anti de Sitter
space

$\bullet$ Quantization and entropy of the surface of a black hole.

$\bullet$ Collapse in diverse dimensions.

$\bullet$ Scalar tensor (or other) gravity theories and collapse

$\bullet$ Collapse with a cosmological constant included.

$\bullet$ Particle production in collapse.

$\bullet$ Role of pressure (radial or tangential) in collapse.

$\bullet$ Stability of stars against collapse due to rotation.

$\bullet$ Gravitational radiation from (non-spherical) collapse.

\section{Acknowledgement}

This work was supported by NASA grant 8-39225 to Gravity Probe B,
and by the US Department of Energy under Contract No.\
DE-AC02-76SF00515. We thank the members of the Gravity Probe B
theory group for many critical and stimulating discussions, in
particular Francis Everitt, Robert Wagoner, and Alex Silbergleit.

\end{document}